\documentclass[%
 reprint,
superscriptaddress,
 amsmath,amssymb,
 aps,
prl,
]{revtex4-2}

\usepackage{graphicx}% Include figure files
\usepackage{dcolumn}% Align table columns on decimal point
\usepackage{bm}% bold math
\usepackage{xcolor}
\usepackage{physics}
\usepackage{quantikz}
\usepackage{amsthm}
\usepackage{hyperref}
\usepackage{bm}
\usepackage{MnSymbol}
\hypersetup{
    colorlinks=true,
    linkcolor=blue!20!black!80,
    filecolor=magenta,      
    urlcolor=cyan,
    }

\newcommand{\E}{\mathcal{E}}
\renewcommand{\O}{\mathcal{O}}
\newcommand{\U}{\mathcal{U}}

\newtheorem{theorem}{Theorem}

\newtheorem{corollary}{Corollary}
\newtheorem{conjecture}{Conjecture}

\begin{document}

% \preprint{APS/123-QED}

\title{Error mitigation for partially error-corrected quantum computers}% \\via quantum circuits for convex combinations of channels}%

\author{Ben DalFavero}
 \affiliation{Department of Computational Mathematics, Science, and Engineering, Michigan State University, East Lansing, MI 48823, USA}
\affiliation{Center for Quantum Computing, Science, and Engineering, Michigan State University, East Lansing, MI 48823, USA}

\author{Ryan LaRose} \thanks{Corresponding author:  \href{rmlarose@msu.edu}{rmlarose@msu.edu}}
\affiliation{Department of Computational Mathematics, Science, and Engineering, Michigan State University, East Lansing, MI 48823, USA}
\affiliation{Department of Electrical and Computer Engineering, Michigan State University, East Lansing, MI 48823, USA}
\affiliation{Department of Physics and Astronomy, Michigan State University, East Lansing, MI 48823, USA}
\affiliation{Center for Quantum Computing, Science, and Engineering, Michigan State University, East Lansing, MI 48823, USA}

\begin{abstract}
We present a method for quantum error mitigation on partially error-corrected quantum computers --- i.e., computers with some logical qubits and some noisy qubits. Our method is inspired by the error cancellation method and is implemented via a circuit for convex combinations of channels which we introduce in this work. We show how logical ancilla qubits can arbitrarily reduce the sampling complexity of error cancellation in a continuous space-time tradeoff, in the limiting case achieving $O(1)$ sample complexity which circumvents lower bounds for sample complexity with all known error mitigation techniques. This comes at the cost of exponential circuit depth, however, and leads us to conjecture that any error mitigation protocol with (sub-)polynomial sample complexity requires exponential time and/or space, even when logical qubits are utilized as a resource. We anticipate additional applications for our quantum circuits to implement convex combinations of channels, and to this end we discuss one application in simulating open quantum systems, showing an order of magnitude reduction in gate counts relative to current state-of-the-art methods for a canonical problem.
\end{abstract}

\maketitle

\textit{Introduction} --- 
Noise in quantum computers necessitates circuit-level methods, conventionally known as quantum error correction and quantum error mitigation, to produce accurate and reliable results. Quantum error correction (QEC) is the holy grail and has seen significant recent progress~\cite{sivak2023real, acharya2023suppresing, bluvstein2024logical, bravyi2024high, acharya2025scaling}, but the relatively high overhead required for QEC has motivated lower overhead techniques known as quantum error mitigation (QEM) in recent years~\cite{Cai_2023, mitiq_2022, cirstoiu2023volumetric}. While QEC and QEM are often viewed as two distinct strategies for two distinct frameworks of quantum computing --- namely, noisy intermediate-scale quantum (NISQ)~\cite{Preskill_2018} computing and fault-tolerant quantum computing (FTQC) --- the transition from NISQ to FTQC is unlikely to be a sudden phase transition. Rather, it will likely be a gradual transition from noisy to logical qubits, with future devices capable of correcting errors on some but not all qubits. To this end, in this work we consider the setting of \textit{partially error-corrected} quantum computers with some logical (perfect) qubits and some physical (noisy) qubits, and we develop a method for executing accurate quantum computations. Our method is inspired by a QEM technique known as probabilistic error cancellation, and works via a circuit construction for implementing convex combinations of channels that we introduce in this work. This circuit construction may be of independent interest and find use in other applications --- for example we also discuss the application of simulating open quantum systems.

The QEM technique known as probabilistic error cancellation (PEC), and sometimes the quasi-probability distribution (QPD), method~\cite{temme_error_2017} has received considerable attention~\cite{Berg_Minev_Kandala_Temme_2023,mitiq_2022,Piveteau_Sutter_Bravyi_Gambetta_Temme_2021,Russo_Mari_Shammah_LaRose_Zeng_2023,Strikis_Qin_Chen_Benjamin_Li_2021} due to its relatively simple procedure as well as its ability to produce exact results, assuming noisy operations can be perfectly characterized. Like several other error mitigation techniques, however, the time to implement the method grows exponentially in some parameter~\cite{Takagi_Endo_Minagawa_Gu_2022}. In particular, the time complexity (number of circuits needed to be executed) of PEC scales as $\Gamma^L$ where $\Gamma$ is the overall negativity and $L$ is the number or operations in the circuit. (More detail below.) Because of this, several authors have sought methods to lower the negativity $\Gamma$ in recent literature~\cite{Mari_Shammah_Zeng_2021, Piveteau_Sutter_Woerner_2022, Hsieh_Tsai_Lu_Li_2024}, either by finding optimal unitary representations, combining PEC with other techniques, or allowing for local operations and classical communication (LOCC). 

In the setting of partially error-corrected computers, we show that our method is able to reduce the negativity (equivalently, the sample complexity) arbitrarily. While this reduction in sample complexity comes at the cost of space complexity --- namely additional qubits and additional operations --- we find this an interesting theoretical result because it bypasses lower bounds on sample complexity required for error mitigation~\cite{Takagi_Endo_Minagawa_Gu_2022, Takagi_Tajima_Gu_2023, eisert2024exponentially}. In particular, Theorem 1 of~\cite{eisert2024exponentially} shows that any (weak) quantum error mitigation algorithm requires a number of samples proportional to $p^{ - \Omega (n D )}$ where $p$ is the local depolarizing noise rate, $n$ is the number of qubits, and $D$ is the depth of the circuit. In contrast, in the limiting case of our method the sample complexity is $O(1)$ on partially error-corrected quantum computers.

\textit{Preliminaries} --- We begin, as in PEC, by assuming perfect knowledge of the noisy implementable operations $\{ \O_\alpha \}_{\alpha = 1}^{M}$ on a given quantum computer. We further assume that these operations form a basis such that any unitary channel $\U$ can be written
\begin{equation} \label{eqn:pec-single-unitary}
    \U = \sum_\alpha c_\alpha \O_\alpha = \gamma \sum_\alpha \sigma_\alpha p_\alpha \O_\alpha .
\end{equation}
Here, $c_\alpha$ are real coefficients, and in the second equality we have defined $\sigma_\alpha := \text{sgn}(c_\alpha)$ and $p_\alpha := |c_\alpha| / \gamma$ with $\gamma := \sum_\alpha |c_\alpha|$. This allows us to view the equality as a quasi-probability distribution with negativity $\gamma$. For a product of $L$ unitaries $\mathcal{C} := \mathcal{U}_L \ \cdots \ \U_1$ --- i.e., a depth $L$ quantum circuit --- we can repeatedly apply~\eqref{eqn:pec-single-unitary} to obtain
\begin{equation} \label{eqn:error-cancellation-circuit}
    \mathcal{C} = \Gamma \sum_{\bm{\alpha}} \sigma_{\bm{\alpha}} p_{\bm{\alpha}} O_{\bm{\alpha}} 
\end{equation}
where now the bold-faced index $\bm{\alpha}$ runs over the tuple $\bm{\alpha} := (\alpha_1, ..., \alpha_L)$ and the overall negativity $\Gamma := \gamma^{[1]} \cdots \gamma^{[L]}$ is the product of negativities $\gamma^{[i]}$ for each unitary $\U_i$. For an observable $A = A^\dagger$ and initial state $\rho$, we can now write
\begin{equation} \label{eqn:error-cancellation-expectation-value}
    \langle A \rangle \equiv \Tr [ \mathcal{C}(\rho) A ] =  \Gamma \sum_{\bm{\alpha}} \sigma_{\bm{\alpha}} p_{\bm{\alpha}} \Tr[ O_{\bm{\alpha}} (\rho) A ] .
\end{equation}
We emphasize this expression is exact and allows us to write an ideal expectation value as a linear combination of noisy expectation values. Let us refer to this as error cancellation. The technique of PEC is to sample $\O_{\bm{\alpha}}$ according to $p_{\bm{\alpha}}$ and compute
\begin{equation}
    \langle A \rangle_{\text{PEC}}(N) := \frac{1}{N} \sum_{i = 1}^{N} \sigma_{\bm{\alpha}} \Tr[ O_{\bm{\alpha}} (\rho) A ]  .
\end{equation}
Letting $\langle A \rangle$ denote the ideal (noiseless) expectation value, to achieve $\left| \langle A \rangle_{\text{PEC}}(N) - \langle A \rangle \right| \le \delta $, we execute $N = O(\Gamma ^ 2 / \delta ^ 2)$ quantum circuits~\cite{temme_error_2017}. Letting $\gamma := \max_{1 \le i \le L} \gamma_i$, we have that $\Gamma  = O( \gamma^L )$ and thus $N = O( \gamma^{2 L} / \delta ^ 2)$, showing the importance of minimizing $\gamma$ (provided that $L$ is fixed by the quantum circuit of interest) for practical applications. Ref.~\cite{Mari_Shammah_Zeng_2021} explored reducing $\Gamma$ by extending the set of implementable operations $\{ \O_{\bm{\alpha}} \}_{\bm{\alpha}}$ via noise scaling; Ref.~\cite{Piveteau_Sutter_Woerner_2022} explored reducing $\Gamma$ by allowing for approximation error and introducing an algorithm for choosing the quasi-probability distribution in a noise-aware manner; and Ref.~\cite{Hsieh_Tsai_Lu_Li_2024} explored reducing $\Gamma$ by grouping unitaries to decrease the total number of operations $L$ in the circuit. 

\textit{Results} --- We write~\eqref{eqn:error-cancellation-expectation-value} as
\begin{widetext}
\begin{equation} \label{eqn:two-shot-error-cancellation}
    \langle A \rangle = \Gamma^{[+]} \Tr\left[ \left( \sum_{\substack{\bm{\alpha} \\ \sigma_{\bm{\alpha}} = +1}} p_{\bm{\alpha}} \O_{\bm{\alpha}} (\rho) \right) A \right] - \Gamma^{[-]} \Tr\left[ \left( \sum_{\substack{\bm{\alpha} \\ \sigma_{\bm{\alpha}} = -1}} p_{\bm{\alpha}} \O_{\bm{\alpha}} (\rho) \right) A \right] .
\end{equation}
\end{widetext}
Here, we have grouped terms according to their signs $\sigma_{\bm{\alpha}} = \pm 1$ and defined coefficients $\Gamma^{[\pm]} := \prod_{\substack{i = 1 \\ \sigma_i = \pm 1}}^{L} \gamma^{[i]}$. Note that each of these two terms represents an expectation value computed with respect to a convex combination of channels $\sum_\alpha p_\alpha \O_\alpha$. By developing quantum circuits for convex combination of channels, which utilize logical ancilla qubits, we show how to evaluate expressions like~\eqref{eqn:two-shot-error-cancellation} in partially error-corrected quantum computers to perform error cancellation. In particular, we show how logical ancilla qubits can reduce the negativity of error cancellation, as stated in the following theorem.

\begin{theorem}[Error cancellation on partially error-corrected quantum computers] \label{thm:main}
    Let $\mathcal{C} := \mathcal{U}_L \ \cdots \ \U_1$ be an $n$-qubit, depth $L$ quantum circuit preparing the state $\rho$, and let $A$ be an observable. Provided $O(k)$ logical qubits, there exists an error cancellation protocol to estimate $\Tr [ \rho A]$ with negativity $O(\gamma^{L - k})$, where $\gamma = \max_{1 \le i \le L} \gamma(U_i)$ and $\gamma(\U_i) := \sum_\alpha |c_\alpha|$ is the negativity of the $i$th unitary (see \eqref{eqn:pec-single-unitary}). Equivalently, there exists an error cancellation protocol using $O(k)$ logical qubits which estimates $\Tr[ \rho A]$ to accuracy $\delta$ with sample complexity $O(\gamma ^ {2(L - k)} / \delta^2)$. 
\end{theorem}

This result provides one perspective as to how QEM and QEC can fit together in the transition from NISQ to FTQC. The power of each additional logical qubit in this setting is that it can be used to reduce the negativity of PEC --- as mentioned a result sought by several authors~\cite{Mari_Shammah_Zeng_2021,Piveteau_Sutter_Woerner_2022,Hsieh_Tsai_Lu_Li_2024}. 

Recall that recent work (namely, Theorem 1 of~\cite{eisert2024exponentially}) has shown that any (weak) quantum error mitigation algorithm requires a number of samples proportional to $p^{ - \Omega (n D )}$ where $p$ is the local depolarizing noise rate, $n$ is the number of qubits, and $D$ is the depth of the circuit. 
In the extreme case of using $k = O(L)$ logical qubits, we see from Theorem~\ref{thm:main} that the negativity becomes $O(1)$ and thus the sample complexity becomes $O(1 / \delta^2)$ for target accuracy $\delta$. In principle this bypasses the exponential sample complexity lower bound via the use of logical ancilla qubits, however the method is not practical due to two reasons. First, we must require that $n > L$, for if we had enough logical qubits there would be no need to perform error cancellation. Second, as will be shown in the method, we require  (partially error-corrected) circuits of depth $O(\exp L)$. For future reference we state this as a corollary:

\begin{corollary}[Constant-sample error cancellation on partially error-corrected quantum computers, at the cost of exponential depth] \label{thm:constant-sample-error-cancellation}
    Let $\mathcal{C} := \mathcal{U}_L \ \cdots \ \U_1$ be an $n$-qubit, depth $L$ quantum circuit preparing the state $\rho$, and let $A$ be a Hermitian operator. Provided $O(L)$ logical ancilla qubits and $n$ noisy qubits, there exists a protocol to estimate $\Tr [\rho A]$ to accuracy $\delta$ using $O(1 / \delta ^2)$ samples and two circuits of depth $O(\exp L)$.
\end{corollary}

Although this result is not practical in its current form, it is notable that it overcomes the best bounds on sample complexity for error mitigation. Notably, Ref.~\cite{Cai_2023} reviews the main quantum error mitigation techniques which have been developed thus far, and in Table IV enumerates the qubit overhead, circuit runtime overhead, and sampling overhead for them. For the techniques of probabilistic error cancellation, Richardson extrapolation, symmetry verification, virtual distillation, and echo verification, one can see that the sampling overheads all grow exponentially in some parameter, typically the error rate of the computer and/or size of the problem. The difference between our technique is the introduction of \textit{logical} qubit overhead to achieve constant sample complexity in the limiting case. While the exponential depth makes this method impractical, we conjecture (see \textit{Conclusions}) that this is not a deficiency of our approach but rather a fundamental feature of error mitigation. In short, we conjecture there is ``no free lunch'' in error mitigation even when using additional (logical) qubits.

Besides this corollary, we believe that Theorem~\ref{thm:main} will provide the most practical applications of our procedure to reduce the sample complexity of error cancellation using an intermediate number $1 < k < L$ of logical qubits. We now begin proving these results, beginning with our quantum circuit for implementing convex combinations of channels which underlies both.

\textit{Methods: Quantum circuits for convex combinations of channels} --- 
Various authors have considered quantum circuits for implementing (single) quantum channels~\cite{Schlimgen_Head-Marsden_Sager_Narang_Mazziotti_2021,Schlimgen_Head-Marsden_Sager-Smith_Narang_Mazziotti_2022,digital_convex2024,Hu_Xia_Kais_2020}, a technique which has manifold applications in the evolution of open quantum systems~\cite{Breuer_Petruccione_2002}. A limited case of combinations of channels is presented in~\cite{digital_convex2024} where linear combinations of single-qubit depolarizing channels are taken. Our work generalizes this to arbitrary convex combinations of quantum channels.

Note that the linear combination of channels is not necessarily a quantum channel. For example, the single-qubit map $\E (\rho) := \E_0 (\rho) - \E_1 (\rho)$
%
% \begin{equation}
%     \E (\rho) := \E_0 (\rho) - \E_1 (\rho)
% \end{equation}
%
where $\E_\alpha := \rho \mapsto |\alpha\rangle \langle \alpha|$ is not (completely) positive.  On the other hand, if $\E_\alpha$ are completely positive and trace-preserving (CPTP) maps and $p_\alpha \ge 0$ are real coefficients which sum to unity, then $\E := \sum_\alpha p_\alpha \E_\alpha$ is a CPTP map. Indeed, preservation of trace follows from $\Tr [ \E ] = \sum_\alpha p_\alpha \Tr [ \E_\alpha ] = \sum_\alpha p_\alpha = 1$, and (complete) positivity follows from linearity with (complete) positivity of each $\E_i$. % Thus, while special cases may arise for certain linear combinations of channels, in this work we focus on the general case of convex combinations of channels.

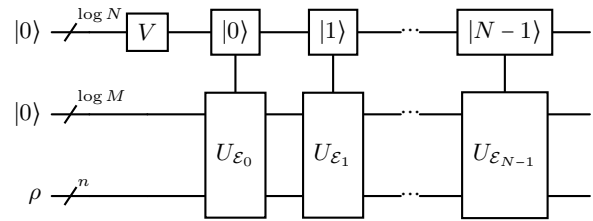
\begin{figure}
    \begin{quantikz}
        \lstick{$\ket{0}$}      & \qwbundle{\log N}        & \gate{V}  & \gate{\ket{0}}\wire[d]{q} & \gate{\ket{1}}\wire[d]{q} & \cdots & \gate{\ket{N-1}}\wire[d]{q}   & \\
        \lstick{$\ket{0}$}   & \qwbundle{\log M}              &           & \gate[2]{U_{\E_0}}        & \gate[2]{U_{\E_1}}        & \cdots & \gate[2]{U_{\E_{N-1}}}        & \\
        \lstick{$\rho $} & \qwbundle{n}       &           & \linethrough              & \linethrough              & \cdots & \linethrough                  & \linethrough  \\
    \end{quantikz}
    \caption{A quantum circuit to implement the convex combination of channels~\eqref{eqn:ccc}. The procedure is similar to the LCU technique with unitary dilations $U_{\E_\alpha}$ for each channel $\E_\alpha$, however there is no post-selection and the preparation is deterministic. The unitary $V$ prepares the probability distribution $V|0\rangle = \sum_\alpha \sqrt{p_\alpha} |\alpha\rangle$ and gates labeled $|\alpha\rangle$ on the top register represent operations controlled on the state $|\alpha\rangle$. The second register used for channel dilations requires $\log M$ qubits, where $M = \max_\alpha M_\alpha$ and $M_\alpha$ is the number of Kraus operators in channel $\E_\alpha$. At the end of the circuit, the bottom register has the convex combination of channels applied to the input state $\rho$, i.e. $\sum_\alpha p_\alpha \E_\alpha ( \rho )$, as proved in the main text.}
    \label{fig:convex-combination-of-channels}
\end{figure}

The quantum circuit in Fig.~\ref{fig:convex-combination-of-channels} proves the following:

\begin{theorem}[Quantum circuit to implement a convex combination of channels]
    For an $n$-qubit state $\rho$, there exists a quantum circuit (Fig.~\ref{fig:convex-combination-of-channels}) to implement
    \begin{equation} \label{eqn:ccc}
        \rho \mapsto \sum_{\alpha = 1}^{N} p_\alpha \E_\alpha(\rho) 
    \end{equation}
    where $p_\alpha \ge 0$ and $\E_\alpha$ are CPTP maps. This circuit uses $n + \log N + \log M$ qubits and has depth $|V| + N$, where $|V|$ is the depth of the unitary $V$ which prepares the coefficient distribution $p_\alpha$, and $M$ is the maximum number of Kraus operators in the channels $\E_\alpha$. The quantum circuit succeeds deterministically, implementing the map~\eqref{eqn:ccc} with probability one.
\end{theorem}

Before proving this result, we note that the technique is similar to the linear combinations of unitaries (LCU)~\cite{Childs_Wiebe_2012,Low_Chuang_2019,Lin_2022} technique, but succeeds deterministically. In particular, the LCU technique implements $|\psi\rangle \mapsto \sum_{\alpha = 1}^{N} c_\alpha U_\alpha |\psi\rangle$ by block encoding and post-selection. Here, a unitary $V$ is used to prepare the (normalized) coefficients $\sum_\alpha \sqrt{c_\alpha} |\alpha\rangle$ from the $|0\rangle$ state on $\log N$ qubits. Subsequently, controlled $U_\alpha$ operations are used to prepare $\sum_\alpha c_\alpha |\alpha\rangle U_\alpha |\psi\rangle$. The coefficient register is then uncomputed such that post-selecting this register on $|0\rangle$ prepares $\sum_\alpha c_\alpha U_\alpha |\psi\rangle$. Because of this, LCU succeeds with a probability that depends on measuring $|0\rangle$ in the coefficient register. This probability depends on the coefficients $c_\alpha$, but can be amplified via amplitude amplification~\cite{berry2014exponential}. In contrast, our construction succeeds deterministically.

\begin{proof}
    It is well known that any CPTP map $\E$ can be written as a unitary acting on a larger Hilbert space followed by a partial trace~\cite{Nielsen_Chuang_2010}. For example, the Stinespring dilation of a channel $\E (\rho) := \sum_{i = 1}^{M} K_i \rho K_i^\dagger$ is given by the unitary $U_\E$ such that
    \begin{equation}
        \langle a, i | U_\E | b, 0 \rangle  = \langle a | K_i | b \rangle .
    \end{equation}
    Here, $|a\rangle$ and $|b\rangle$ index basis states for the system register $H_S$ and $|i\rangle$ indexes basis states for the environment register $H_E$ which has maximum dimension $\dim H_E \le \dim^2 H_S$. Other unitary dilations include the Sz.-Nagy dilation~\cite{Hu_Xia_Kais_2020} as well as symmetric and antisymmetric exponentials as in~\cite{Schlimgen_Head-Marsden_Sager_Narang_Mazziotti_2021}.

    To implement the convex combination of channels $\sum_\alpha p_\alpha \E_\alpha$ where each channel has Kraus operators $K_{1, \alpha}, ..., K_{M_\alpha, \alpha}$, we apply a unitary dilation to each channel $\E_\alpha \mapsto U_{\E_\alpha}$ and proceed as in Fig.~\ref{fig:convex-combination-of-channels}. To see this circuit implements the desired convex combination of channels, assume for simplicity that $\rho = |\psi\rangle \psi|$ is a pure state. Then, the final state is
    \begin{equation}
        \ket{\Psi} = \sum_{\alpha=1}^N \sum_{j=1}^{M_\alpha} \sqrt{p_\alpha} \ket{\alpha}\ket{j} K_{j, \alpha}\ket{\psi}
    \end{equation}
    Taking the partial trace over the first (top) and second (middle) registers gives the reduced density matrix
    \begin{equation}
        \rho = \text{Tr}_{1,2} [\ket{\Psi} \bra{\Psi}] = \sum_{\alpha=1}^N \sum_{j=1}^{N_\alpha} p_\alpha K_{j, \alpha} \ket{\psi} \bra{\psi} K_{j, \alpha}^\dagger 
    \end{equation}
    by virtue of the states \(\ket{\alpha}\) for the first register and \(\ket{j}\) for the second register being orthonormal. Extension from pure state input to mixed state input follows by linearity. 
\end{proof}

Note that the controlled operations enable us to reuse the same ancillary qubits for each channel dilation. It is also possible to introduce new ancilla qubits between each dilation, either through measure-and-reset of the previous ancilla or measurement of the previous ancillae and introduction of new ancillae.

\textit{Application 1: Error cancellation} ---
It is now straightforward to prove Theorem~\ref{thm:main} and Corollary~\ref{thm:constant-sample-error-cancellation}. The two terms of the error cancellation method~\eqref{eqn:two-shot-error-cancellation} are both convex combinations of channels, and so can be evaluated exactly via two applications of the circuit in Fig.~\ref{fig:convex-combination-of-channels}. The total number of terms in~\eqref{eqn:two-shot-error-cancellation} is $N = O(\exp L)$, and therefore we require $O(L)$ logical qubits in the top register of Fig.~\ref{fig:convex-combination-of-channels}. Note that the second register of $\log M$ qubits is not necessary for error cancellation because we assume the operations themselves are noisy channels. When controlling operations on logical qubits (register $C$) acting on the target system (register $S$), we assume that the noise acts as
\begin{equation}
    \rho_{CS} \mapsto \sum_{\alpha} (I_C \otimes K_\alpha) \rho_{CS} (I_C \otimes K_\alpha)^\dagger .
\end{equation}
This assumption is natural since any errors which occur on logical qubits, or which occur on physical qubits and spread to logical qubits, will be able to be corrected on the logical qubits. Under this assumption, it is sufficient to characterize the noisy implementable operations on the noisy system, and use these characterizations in the error cancellation procedure. The only subtlety is that a controlled implementable operation may not be directly implementable. For example, on a computer with gateset $\{U, \text{CNOT}\}$ with $U$ being any single-qubit unitary, $CZ$ is not a directly implementable. This subtle issue can be resolved by simply compiling the operation into a sequence implementable gates, then characterizing this sequence. For example, we would compile $CZ$ to $(I \otimes H) \text{CNOT} (I \otimes H)$, then characterize this sequence. This completes the proof of Corollary~\ref{thm:constant-sample-error-cancellation}. 

The proof of Theorem~\ref{thm:main} follows the same line of reasoning, just applying the circuit for a convex combination of channels to a subset of operations and sampling from the remaining operations as in PEC. For example, let $\mathcal{C} = \U_2 \U_1$ with $\U_i$ as in~\eqref{eqn:pec-single-unitary}.
% %
% \begin{equation}
%     \mathcal{U}_i = \gamma^{[i]} \sum_\alpha \sigma_\alpha^{[i]} p_\alpha^{[i]} \O_\alpha^{[i]} .
% \end{equation}
% %
Rather than implement both unitaries via convex combinations of channels, we implement only one of them via convex combinations of channels, and sample operations from the other. That is, we can write
%
% \begin{widetext}
    \begin{align*} \label{eqn:ccc-pec-example}
            &\Tr [ \rho A] = \gamma^{[2]} \sum_\alpha \sigma_\alpha^{[2]}  p_{\alpha}^{[2]} \times \\ & ( \gamma^{[1, +]} \Tr[ (\O^{[1, +]} \circ \O_\alpha)(\rho) A]  - \gamma^{[1, -]} \Tr[ (\O^{[1, -]} \circ \O_\alpha)(\rho) A] )
    \end{align*}
% \end{widetext}
%
where $\O^{[1, \pm]} := \sum_{\substack{\alpha \\ \sigma_\alpha = \pm 1}} p_\alpha^{[1]} O_\alpha$ is implemented via Fig.~\ref{fig:convex-combination-of-channels} and the remaining term is implemented by sampling $p_\alpha^{[2]}$ as in PEC. Letting $\gamma^{[1]} = \gamma^{[2]} = \gamma$ as before, we see directly from this expression that negativity of PEC is reduced by a factor of $\gamma$. Note that in this case $O(1)$ logical qubits are required to implement the circuit in Fig.~\ref{fig:convex-combination-of-channels}.  The generalization to $O(k)$ logical qubits follows directly.

% error mitigation techniques such as virtual distillation~\cite{virtual_distillilation} which require multiple copies of the state and (controlled) multi-qubit operations to perform SWAP tests.

\textit{Application 2: Simulating open quantum systems} ---
While we have focused on the application of error cancellation, there are likely additional applications of our circuit for convex combinations of channels. Here we describe one, namely the simulation of open quantum systems.
In the study of open quantum systems, one commonly uses the Born and Markov approximations to simplify model-building. Any physics not in agreement with the Born and Markov approximations is commonly said to be non-Markovian. The general treatment of non-Markovian open systems often involves integrodifferential master equations~\cite{zwanzig_ensemble1960,nakajima_on_quantum_theory}, approximate time-convolutionless master equations~\cite{CHRUSCINSKI2016399}, or approaches which emulate the bath using a smaller, ancillary system \cite{pseudomode,kretschmer_collisional}. Recent studies have found special cases where non-Markovian dynamics can come from convex combinations of Markovian channels \cite{JAGADISH2020126907,non_markovianity_degree,assessing_non_markovian}. Such constructions allow the implementation of non-Markovian dynamics without the need to explicitly simulate the environment, or other exponential overhead. Additionally, Ref.~\cite{digital_convex} implements convex mixtures of single-qubit Pauli channels to simulate both Markovian and non-Markovian dynamics. Finally, Ref.~\cite{JAGADISH2020126907} discusses the role of CP divisibility in determining whether a channel is Markovian or non-Markovian. 
% If a channel cannot be decomposed into a composition of two CPTP channels at all times, then the evolution of the system is said to be non-Markovian. This kind of CP-indivisibility is strongly associated with the negative decay rates that show up in time-local master equations. 
Our circuit construction for implementing convex combinations of channels thus presents an avenue for both simulating (non-)Markovian dynamics and to test if a given evolution expressed as a convex combination of channels is (non-)Markovian. %  by combining our circuits with tests for non-Markovianity, e.g. CP divisibility or the RHP criterion~[].

To illustrate one application and the advantage of our approach, consider the Lindblad equation
\begin{equation}
    \dot{\rho} = -i [H, \rho] + \sum_k \Gamma_k \left( L_k \rho L_k^\dagger - \frac{1}{2} \{ L_k, L_k^\dagger, \rho \} \right),
\end{equation}
a master equation for Markovian open quantum systems commonly employed in atomic, molecular, and optical physics, among other applications. Recent work~\cite{david_faster_2024} has identified a technique for solving the Lindblad equation on a quantum computer that casts the problem as a convex combination of channels, similar to the quantum stochastic drift protocol used for unitary evolution~\cite{campbell_random_2019}. While this convex combination of channels can be implemented by classically sampling the channels, the work by David et al.~\cite{david_faster_2024} proposes using the quantum forking protocol~\cite{park_parallel_2019} to implement the combination with a single quantum circuit. The quantum forking protocol requires multiple registers, one for each channel, each of which is the size of the system the channels are being applied to. However, our construction (Fig.~\ref{fig:convex-combination-of-channels}) uses the same register to implement an arbitrary convex combination of channels. Additionally, quantum forking requires that a deep network of controlled swap gates be applied to the system registers before and after the channels are applied. This must be done for each time step in the solution to Lindblad equation. As a tradeoff, our method requires the channels to be controlled. To get a sense of the advantage of our approach, we considered a simple canonical problem of a two level atom subject to damped Rabi flopping. This system is described by the Lindbladian
\begin{equation}
    \label{eq:liouvillian}
    \dot{\rho} = -i [\rho, H] + \gamma (\sigma^- \rho \sigma^+ + \frac{1}{2}\{\sigma^- \sigma^+, \rho\}),
\end{equation}
where $H = \omega_0 \sigma_Z + \Omega \sigma_X$, $\sigma_X$ and $\sigma_Z$ are the Pauli matrices, and $\sigma^- = \ket{g}\bra{e}$, $\sigma^+ = \ket{e}\bra{g}$. Compared to the quantum forking method, our circuit requires roughly an order of magnitude fewer gates, as shown in Table~\ref{tab:damped-rabi-gate-counts}. More details on this comparison are given in the Appendix.

\begin{table}[]
    \centering
    \begin{tabular}{|c|c|c|c|}
        \hline
        Method & Qubits & Two-qubit gates & Circuit depth \\ \hline
        Refs.~\cite{david_faster_2024,park_parallel_2019}  & 8 & 1106 & 2197 \\
        This work & 4 & 126 & 253 \\
        \hline
    \end{tabular}
    \caption{Qubits, two-qubit gates, and circuit depths for simulating damped Rabi flopping~\eqref{eq:liouvillian} with the method of Ref.~\cite{david_faster_2024, park_parallel_2019} and with our method for implementing convex combinations of channels (Fig.~\ref{fig:convex-combination-of-channels}).}
    \label{tab:damped-rabi-gate-counts}
\end{table}

\textit{Conclusion} --- In this Letter, we have introduced a method for implementing error cancellation on partially error-corrected quantum computers, showing how logical qubits can reduce the sample complexity of probabilistic error cancellation. In the theory of error mitigation, our work bypasses previous results showing any error mitigation algorithm must consume an exponential number of samples. Indeed, we have discussed how the extreme limit of our method allows for error cancellation with constant sample complexity. While this requires exponential space (gate depth) overhead to do, we find it unlikely that any error mitigation procedure will achieve ``a free lunch'' with polynomial time, space, and sample complexity. Indeed, as shown in Table IV of Ref.~\cite{Cai_2023}, all known error mitigation techniques with constant or polynomial space and time resources require exponential sample complexity. In light of these results and our results in Theorem~\ref{thm:constant-sample-error-cancellation} and Corollary~\ref{thm:constant-sample-error-cancellation}, we conjecture:
\begin{conjecture}
    Any error mitigation technique achieving (sub-)polynomial sample complexity must require exponential space and/or time complexity, even when logical qubits are utilized as a resource.
\end{conjecture}
Nonetheless, we believe Theorem~\ref{thm:main} will provide a useful error mitigation protocol as we begin to have quantum computers with some logical qubits. This protocol is interesting in that it shows one method in which QEM and QEC can work together, utilizing QEC to reduce the (sampling) overhead of QEM. We emphasize that, when targeting practical applications, known methods to reduce the overhead of error cancellation can be applied for our protocol as well --- e.g., grouping operators to reduce the total number of gates $L$ in the circuit, tailoring device noise by Pauli twirling to simplify unitary representations, dropping terms with small coefficients to produce an approximate result with lower overhead, and even combining the protocol with other methods like zero-noise extrapolation to reduce the negativity further. It may even be possible to perform our method with noisy qubits in place of logical qubits. This would likely require noise tailoring on the top register of Fig.~\ref{fig:convex-combination-of-channels} and careful consideration of how implementable operations in error cancellation get controlled in our circuit for convex combinations of channels. Generally, we hope our work inspires new directions in error suppression in the transition from NISQ computing to FTQC, including both the theory and application of quantum error mitigation and the interplay between QEM and QEC~\cite{aharanov2025on}.

We expect our circuit for implementing convex combinations of channels to find use in other applications, and to that end we have discussed on application in simulating open quantum systems. Even for a simple canonical example, we have seen that our circuit achieves a substantial reduction in both qubits and gate counts relative to current state-of-the-art methods. Our circuit is therefore a promising method for simulating and probing Markovian and non-Markovian evolution, an application with practical and theoretical value. 

\textit{Acknowledgments} --- We acknowledge support from the Wellcome Leap Quantum for Bio Program.

\bibliography{refs.bib}

\appendix

\section{More details on gate counts for Rabi flopping} \label{app:rabi-flopping}

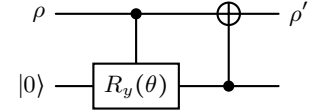
\begin{figure}
    \centering
    \begin{quantikz}
    \lstick{$\rho$} & \ctrl{1}  & \targ{} & \rstick{$\rho'$} \\
    \lstick{$\ket{0}$} & \gate{R_y(\theta)} & \ctrl{-1} & \\
    \end{quantikz}
    \caption{A unitary circuit to implement the amplitude damping channel with input state $\rho$ and output state $\rho'$. The damping parameter is $\beta = \sin^2(\theta)$~\cite{Nielsen_Chuang_2010}.}
    \label{fig:amplitude-damping-circuit}
\end{figure}

Here we provide more details on computing the gate counts shown in Table.~\ref{tab:damped-rabi-gate-counts}. To solve~\eqref{eq:liouvillian}, evolution under the Liouvillian is split into three channels:
\begin{equation*}
\begin{split}
    \mathcal{O}_1[\rho] &= e^{-i \sigma_Z \tau} \rho e^{i \sigma_Z \tau} \\
    \mathcal{O}_2[\rho] &= e^{-i \sigma_X \tau} \rho e^{i \sigma_X \tau} \\
    \mathcal{O}_3[\rho] &= E_0 \rho E_0^\dagger + E_1 \rho E_1^\dagger
\end{split}
\end{equation*}
each of which has the associated probability
\begin{equation*}
    \begin{split}
        p_1 &= \frac{\omega_0}{\lambda} \\
        p_2 &= \frac{\Omega}{\lambda} \\
        p_3 &= \frac{\gamma}{\lambda}
    \end{split}
\end{equation*}
where $\lambda = \omega_0 + \Omega + $and $\tau = \lambda \delta t$, $\delta t$ is the length of the time step, and the Kraus operators $E_0$ and $E_1$ correspond to an amplitude damping channel:
\begin{equation*}
    \label{eq:amplitude-damping-kraus-ops}
    \begin{split}
        E_0 &= 
        \begin{bmatrix}
            1 & 0 \\
            0 & \sqrt{1 - \beta}
        \end{bmatrix} \\
        E_1 &= 
        \begin{bmatrix}
            0 & \sqrt{\beta} \\
            0 & 0
        \end{bmatrix} \\
    \end{split}
\end{equation*}
with $\beta = e^{-\gamma t}$. A circuit to simulate this channel using one ancilla qubit for dilation~\cite{Nielsen_Chuang_2010} is shown in Fig.~\ref{fig:amplitude-damping-circuit}.

Using these channels, we implement both the method of~\cite{david_faster_2024,park_parallel_2019} and our method in Fig.~\ref{fig:convex-combination-of-channels} to simulate the evolution of~\eqref{eq:liouvillian}. The resulting circuits are compiled to a gateset consisting of CNOT gates and arbitrary single-qubit rotations. These gate counts are shown in Table~\ref{tab:damped-rabi-gate-counts}.

\end{document}